\documentclass[9pt]{article}
\usepackage{spconf,amsmath,graphicx}
\usepackage{algorithm,algorithmic}
\usepackage{amsthm,amsmath,amssymb}
\usepackage{color}
\usepackage{graphicx}
\title{Sequential Adaptive Detection for \\In-situ transmission electron microscopy (TEM)}
\name{Y. Cao$^{\star}$ \quad S. Zhu$^{\star}$ \quad Y. Xie$^{\star}$\sthanks{Corresponding author: yao.xie@isye.gatech.edu}  \quad J. Key$^{\dagger}$ \quad J. Kacher$^{\dagger}$\sthanks{The work is sponsored by a Georgia Tech IMAT Seed Grant.} \quad R. R. Unocic$^\ddag$ \quad C. M. Rouleau$^{\ddag}$\sthanks{The TEM corrosion experiments were conducted at the Center for Nanophase Materials Sciences, which is a DOE Office of Science User Facility.}}

\address{$^{\star}$ School of Industrial and Systems Engineering, Georgia Institute of Technology, Atlanta, GA, USA. \\
    $^{\star}$ School of Material Science and Engineering, Georgia Institute of Technology, Atlanta, GA, USA.\\
    $^{\dagger}$Oak Ridge National Laboratory, Oak Ridge, TN, USA. }

\begin{document}
%\ninept
%
\maketitle
\begin{abstract}
We develop new efficient online algorithms for detecting transient sparse signals in TEM video sequences, by adopting the recently developed framework for sequential detection jointly with online convex optimization \cite{cao2017near}. We cast the problem as detecting an unknown sparse mean shift of Gaussian observations, and develop adaptive CUSUM and adaptive SSRS procedures, which are based on likelihood ratio statistics with post-change mean vector being online maximum likelihood estimators with $\ell_1$. We demonstrate the meritorious performance of our algorithms for TEM imaging using real data. 
\end{abstract}
\section{Introduction}
\label{sec:intro}

TEM (Transmission Electron Microscopy) has long been a powerful tool for imaging material structure and characterizing material chemistry. However, the process to resolve structural features is laborious and time intensive, drastically limiting the characterization throughput. Recent advances in electron detector technology and computational capacity have facilitated the development of high-speed data collection with microsecond frame rate acquisition speeds. This advance in TEM technology has enabled new paradigms in data collection.  
Because of this, {\it in-situ} processing of the real-time collected data to detect emerging features become a highly desired property for the new TEM system. Currently, the data are captured real-time but analyzed off-line, limiting the experimentalist's ability to explore in detail regions of interest while at the microscope. 

Sequential change-point detection that can be adaptive to data can revolutionize this process. 
The signal detection task in TEM has two characteristics. First, each observation is a very high-dimensional vector so we need develop an algorithm that can handle a large amount of data sequentially. Second, the change is \textit{sparse} in the sense that among all the parameters only a small proportion of them changes after the unknown change-point.

In this paper, we present a sequential adaptive change detection method for in-situ TEM signal detection. The method is developed by adapting the recent one-sample update based sequential detector in \cite{cao2017near}, by assuming Gaussian observations and the signal being a sparse mean shift to the Gaussian. Our method can precisely control false alarms and can be computed recursively, and thus automate the detection in real-time. We demonstrate meriterous performance of our methods for TEM imaging using real data.

Compared to the classic CUSUM procedure (see, e.g., \cite{tartakovsky2014sequential}), which needs to pre-specify a post-change mean parameter and its performance can can impacted when there is parameter misspecification, our adaptive procedure is more robust since the mean is updated with sequential data. Compared to the classic generalized likelihood ratio (GLR) procedure when the plug-in estimators are exact maximal likelihood estimators (MLE), our method is much faster and memory efficient since our plug-in estimators are computed recursively with one-sample update (thus raw data needs not to be stored) using an online convex optimization algorithm.

%Due to the high vacuum required for TEM, liquid cannot usually be introduced to the system. However, the development of liquid cell TEM holders now allows samples to be exposed to aqueous environments through the use of sealed microfluidic chips \cite{chee2015studying}.

%\noindent{\bf Background of TEM}. 
%%
%Transmission electron microscopy (TEM) is an imaging technique in material science, which uses a series of lenses and apertures to form an image from electrons that have been transmitted through a sample. In bright field mode, shown in Figure \ref{fig:TEM}, the objective aperture is used to block scattered electrons, and the image is formed from the beam of electrons that passes directly through the sample. The image contrast  comes from differences in scattering, which can be caused by changes in sample thickness, phase, or crystal orientation \cite{williams1996transmission,chee2015studying}. We are interested in performing {\it in-situ} analysis of a sequence of TEM images. In particular, we would like to detect in real-time the corrosion pit initiation for understanding when and where they start. This is crucial for relating the material property to corrosion susceptibility.   

\begin{figure}[h!]
\centering
\includegraphics[width=1\linewidth]{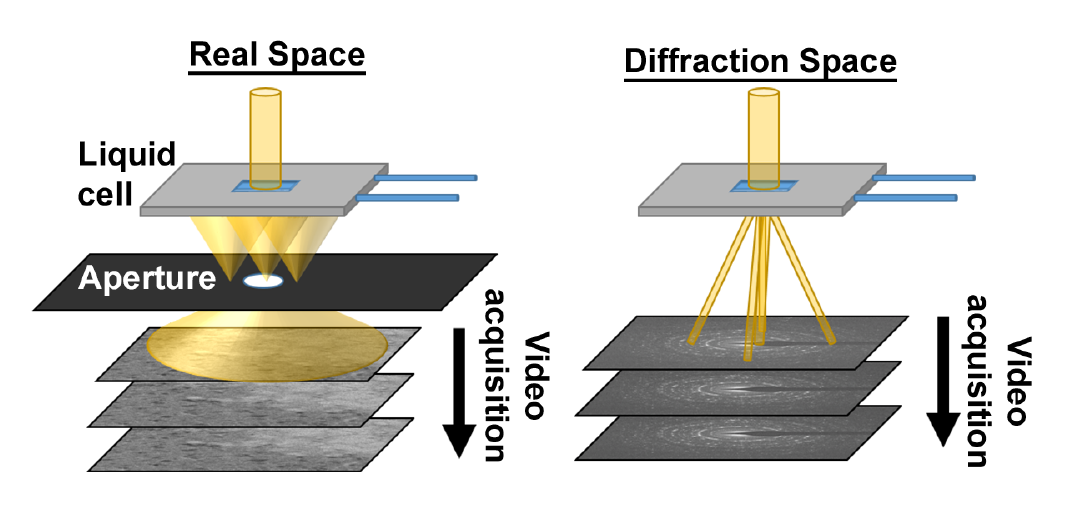}
\caption{Diagrams demonstrating the basic principles of bright field imaging in TEM. TEM can operate in two modes, illustrated in Left Panel: in the real space; Right Panel: in the diffraction space. The real space images can be computed from the diffraction space images. The incident beam of electrons passes through the sample and a lens/aperture system is used to form the image. Our algorithm can be applied to both type of data: the image sequence data and the diffraction pattern sequence data.}
\label{fig:TEM}
\vspace{-0.1in}
\end{figure}

\section{Sequential adaptive detection}

In this section, we present our two sequential adaptive detection algorithms, which are adapted from the one-sample update scheme in \cite{cao2017near}. 
Assume a sequence of $d$-dimensional observations $X_1, X_2$, $\ldots$ which are i.i.d. random variables from a multivariate normal distribution $\mathcal{N}(\theta, I_d)$ with {\it unknown} mean parameter $\theta \in \mathbb{R}^d$. We will estimate $\theta$ online to be adaptive.

Consider the sequential change-point detection problem that the underlying distribution of the data changes from a known state to an unknown state after at an unknown change-point $\nu$. Without loss of generality, we assume that the pre-change mean is an all-zero vector. The post-change mean is unknown and belong a set $\mathcal{A}$ defined as $\mathcal{A} = \{\theta: \|\theta\|_0 \leq s \}$, where $\|\cdot\|_0$ is the number of non-zero entries of $\theta$ and $s$ is a prescribed value to characterize the sparsity. Formally, we consider the following hypothesis test: 
\begin{equation}
\begin{split}
\textsf{H}_0: &~~X_1, X_2, \ldots \overset{\rm i.i.d.}{\sim} \mathcal{N}(0,I_d), \\ 
\textsf{H}_1: &~~X_1, \ldots, X_{\nu} \overset{\rm i.i.d.}{\sim}\mathcal{N}(0,I_d), \\
&~~X_{\nu+1}, X_{\nu+2}, \ldots \overset{\rm i.i.d.}{\sim} \mathcal{N}(\theta,I_d), ~~\theta \in \mathcal{A}.
\end{split}
\label{maintestproblem}
\end{equation}
The goal is to detect the change as quickly as possible after it occurs under the false alarm constraint. 
We will consider likelihood ratio based detection procedures which we call the adaptive CUSUM (ACM), and the adaptive SRRS (ASR) procedures, respectively. 

Now we derive the detection statistics. For each putative change-point location $k$ before the current time $t$, the post-change samples are $\{X_{k}, \ldots, X_t\}$, and the post-change parameter is estimated as 
\begin{equation}
\hat{\theta}_{k,i} = \hat{\theta}_{k,i}(X_k, \ldots, X_{i}), \quad i\geq k.
\label{eq:theta_change}
\end{equation} 
Denote $f_{\theta}$ as the density function for $\mathcal{N}(\theta, I_d)$. The likelihood ratio at time $t$ for a hypothetical change-point location $k$ is given by (initialized with $\hat{\theta}_{k,k-1} = \theta_0$)
\begin{equation}
\Lambda_{k,t} = \prod_{i=k}^t  \frac{f_{\hat{\theta}_{k,i-1}}(X_i)}{f_{0}(X_i)}, 
\label{cumulativestat}
\end{equation}
where $\Lambda_{k, t}$ can be computed recursively since
\[
\Lambda_{k,t} = \Lambda_{k,t-1} \cdot \frac{f_{\hat{\theta}_{k,t-1}}(X_t)}{f_{0}(X_t)}.
\]

Since the change-point location $\nu$ is unknown, due to the maximum likelihood principle, we take the maximum of the statistics over all possible values of $k$. We consider window-limited versions \cite{willsky1976generalized} to avoid infinite memory, by taking the maximum over $k\in [t-w,t]$, where $w$ is a prescribed window size. This leads to the ACM procedure 
\begin{equation}
T_{\rm ACM}(b) = \inf \left\{t\geq 1: \max_{t-w\leq k\leq t} \log \Lambda_{k,t} > b\right\}, 
\label{ACMprocedure}
\end{equation}
where $b$ is a pre-specified threshold.

The Shiryaev-Roberts (SR) procedure replace the maximization over $k$ in (\ref{ACMprocedure}) with summation, which can be justified from a Bayesian prior assumption. By following the same strategy, we obtain the following ASR procedure \cite{lorden2005nonanticipating}:
\begin{equation}
T_{\rm ASR}(b) = \inf \left\{t\geq 1: \log \left(\sum_{k=t-w}^t \Lambda_{k,t} \right)> b \right\},
\label{ASRprocedure}
\end{equation}
where $b$ is a pre-specified threshold. As shown in \cite{cao2017near}, the performance of the ACM and the ASR is very similar. However, the likelihood ratio in (\ref{cumulativestat}) can explode when $d$ is very large. Thus, in practice we prefer to use the ACM procedure to avoid possible numerical issues.

The detection statistic relies on a sequence $\{\hat{\theta}_{k,t}\}$ of estimators constructed using online mirror descent (OMD). The main idea of OMD is that, at each time step, for any $k$, the estimator $\hat{\theta}_{k,t-1}$ is updated using the new sample $X_t$, by balancing the tendency to stay close to the previous estimate against the tendency to move in the direction of the greatest local decrease of the loss function. The advantages of OMD are (1) it allows a simple {\it one-sample update}: the update from $\hat{\theta}_{k,t-1}$ to $\hat{\theta}_{k,t}$ only uses the current sample $X_t$, and the update for the detection statistic has a simple recursive scheme. This is the main difference from the traditional GLR statistic \cite{lai1998information} where each $\hat{\theta}_{k,t}$ is the exact MLE estimated using all the historical samples. (2) OMD is a generic algorithm for solving the online convex optimization (OCO) problem \cite{hazan2016introduction}. %If we set the loss function in OCO framework as the negative log-likelihood then the output of one OCO algorithm is a sequence of estimators that are the approximations of the exact MLEs. 
In \cite{cao2017near}, it is proven that even these approximate MLE schemes have very little statistical efficiency. 

Here, we adapt the general ACM and ASR for exponential family distributions in \cite{cao2017near} to the case when the signal is a sparse Gaussian mean shift, and set the constraint set for the unknown parameter to be $\Gamma = \{\theta: \|\theta\|_1 \leq s \}$ as a convex relaxation of the non-convex set $\mathcal{A}$. Denote $\|\cdot\|_1$ and $\|\cdot\|_2$ as the $\ell_1$ and $\ell_2$ norms in the Euclidean space, respectively. The algorithms are summarized in Algorithm \ref{alg1}. The projection (step 7) onto $\ell_1$ ball can be obtained via simple soft-thresholding \cite{duchi2008efficient}.

\begin{algorithm}[h!]
\caption{Online mirror-descent (OMD) for $\{\hat{\theta}_{k,t}\}$
}\label{alg1}
\begin{algorithmic}[1]

\REQUIRE A sequence of data $X_k, \ldots \in \mathbb{R}^d$; a closed and convex set $\Gamma \subset \mathbb{R}^d$ of the parameters; a decreasing sequence $\{ \eta_t\}_{t\geq 1}$ of strictly positive step-sizes.
%Exponential family specifications $\phi(x), \Phi(x)$ and $f_{\theta}(x)$; initial parameter value $\theta_0$; sequence of data $X_k, \ldots, X_t, \ldots$;  
%\vspace{.1in}
\STATE $\hat{\theta}_{k,k-1} =0, \Lambda_{k,k-1} = 1$. \COMMENT{Initialization}

%\vspace{.1in}
\FORALL{$t = k,k+1,\ldots,$}

\STATE \mbox{Acquire a new observation $X_t$}

%\vspace{.1in}
\STATE \mbox{Compute loss $\ell_t(\hat{\theta}_{k,t-1}) :=  \|\hat{\theta}_{k,t-1}\|_2^2/2 - \hat{\theta}_{k,t-1}^\intercal X_t$}

%\vspace{.1in}
\STATE 
\mbox{Compute $\Lambda_{k,t} = \Lambda_{k,t-1} \times f_{\hat{\theta}_{k,t-1}}(X_t)/f_{0}(X_t)$}
 %\COMMENT{Compute}

%Incur a cost $l_i(\hat{\theta}_{i-1}) = -\log f_{\hat{\theta}_{i-1}}(X_i) = \Phi(\hat{\theta}_{i-1}) - \hat{\theta}_{i-1}^\intercal \phi(X_i)$

%\vspace{.1in}
\STATE $\tilde{\theta}_{k,t} = \hat{\theta}_{k,t-1} - \eta_t(\hat{\theta}_{k,t-1} - X_t)$
\COMMENT{Dual update}

%\vspace{.1in}
%\STATE $\tilde{\theta}_{k,t} = (\nabla \Phi)^{*}(\hat{\mu}_{k,t})$

%\vspace{.1in}
\STATE  $\hat{\theta}_{k,t} = \mathop{\arg\min}_{u \in \Gamma} \| u-\tilde{\theta}_{k,t} \|_2$  \COMMENT{Projected primal update}
%\vspace{.1in}
\ENDFOR
%\vspace{.1in}
\RETURN $\{\hat{\theta}_{k,t}\}_{t\geq 1}$ and $\{\Lambda_{k,t}\}_{t\geq 1}$.

\end{algorithmic}

\end{algorithm}

\section{Results on real-data}

We test our methods one two TEM datasets: one consists of a sequence of real space images, and another one consists of a sequence of diffraction space images. The experimental set ups are the exact same for the two datasets. The only difference is that we change the lens setting to collect diffraction patterns instead of image-space images. We will develop different preprocess steps for these two datasets due to their different characteristics. After the preprocessing, we show that both become detecting a sparse signal in Gaussian noise.

\begin{figure}[h!]
\centering
\includegraphics[width=1.0\linewidth]{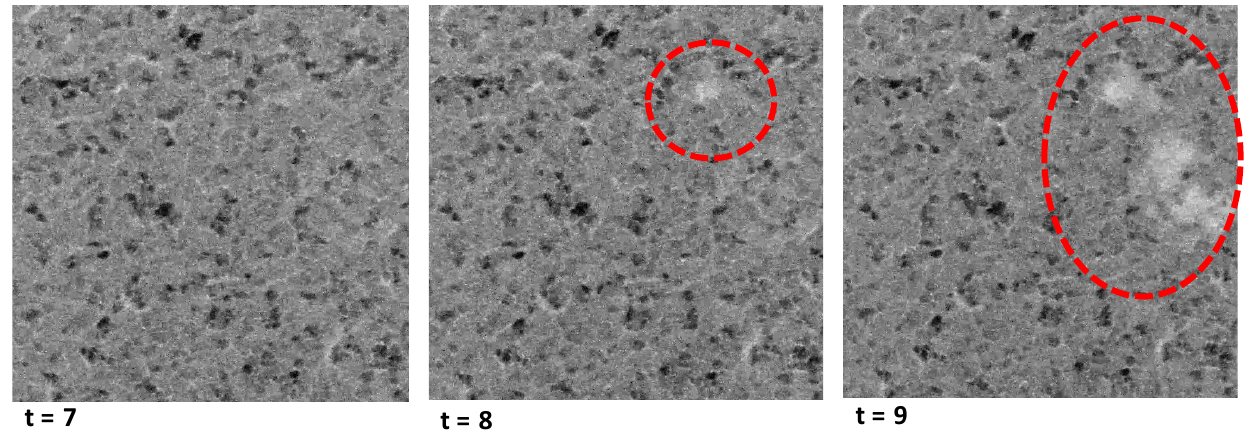}
\caption{A sequence of metal corrosion images captured using bright-field TEM. The time (index for the image in the sequence) is labeled; the corrosion initiates at time t = 8 (marked by the red circle) and develops over time. The corroded area has a higher intensity signal than the rest of the film.}
\label{fig:rust}
\vspace{-0.1in}
\end{figure}

\noindent {\bf Experiment set-up.} The data is a sequence of metal corrosion images captured using bright-field transmission electron microscopy (TEM). The experiment setup is as follows. Iron thin films were sputtered at room temperature onto silicon nitride membranes compatible with an in-situ TEM liquid cell holder. 20 vol\% acetic acid was introduced to the system to initiate corrosion. Imaging was performed using an FEI Titan at 300 kV with a Gatan OneView camera in either real space or diffraction space. The time-resolved diffraction patterns provide information on the formation of corrosion bi-products and the dissolution of crystalline material. For illustration purposes, we first select 23 gray images (2 images per second) in the bright-field TEM image sequence and downsize each image to 308-by-308 pixels. At some time point, corrosion initiates in the image sequence, which is emphasized by the red circle in Figure \ref{fig:rust}.

We apply our ACM and ASR procedures, choosing $w=200$ as the prescribed window size and setting the threshold $b$ for detection procedures by simulation such that the false alarm rate (the average-run-length, ARL, which expected number of observations between two false alarms) is about $10000$. 

\subsection{Detection for real space images}

\noindent{\bf Preprocessing.} 
First, vectorize each image into a vector of dimension $308\times 308 = 94864$.  For each pixel, we take their value in the first $5$ frames as the training data to compute the mean and standard deviation. Then we standardize samples for each pixel by subtracting the mean and dividing the standard deviation. We ignore the correlation between the pixels for this example and it turns out to be a good approximation. 

After the pre-processing, we have a total of 23 such vectors: $X_1, \ldots, X_{23} \in \mathbb{R}^{d}$, with $d=94864$. The pre-change distribution is $\mathcal{N}(0,I_d)$ and the goal is to detect the unknown time $\nu$ at which the underlying distribution changes to $\mathcal{N}(\theta, I_d)$ for some unknown mean $\theta \neq 0$. We assume that on average each pixel has a unit shift after the standardization so we set $s=10^5$ in $\Gamma$ for our methods. We compare our algorithms with the standard multivariate CUSUM procedure \cite{woodall1985multivariate} (the post-change mean parameter is set to be an all-one vector), and the GLR procedure \cite{willsky1976generalized}. 
%The thresholds $b$'s of all procedures are calibrated such that the ARLs for all methods are $10000$. 

\noindent{\bf Results.} $T_{\rm ACM}(b)$ and $T_{\rm ASR}(b)$ with $\Gamma = \{\theta: \| \theta \|_1 \leq 10^5 \}$ both stop at time $t=8$, CUSUM procedure with an all-one post-change mean vector stops at time $t=9$ and GLR procedure stops at time $t=6$. Since we see from Figure \ref{fig:rust} that the change happens at time $t=8$, the detection delays of our methods are $0$ (meaning it only takes one sample to detect the corrosion spot) while that of CUSUM procedure is $1$ (meaning it takes two samples to detect). The GLR procedure raises an false alarm since it stops when there is no change. The possible reason for the GLR raising the false alarm is that the GLR is more easily affected by noise. This also shows that GLR procedure performs better in ideal case such as synthetic signals but may not perform well in practice. 

\subsection{Detection for diffraction space images.}

%\begin{figure}[h!]
%  \begin{subfigure}[b]{0.4\linewidth}
%    \includegraphics[width=\linewidth]{original_figure.pdf}
%    \caption{}
%  \end{subfigure}
%  \hfill
%  \begin{subfigure}[b]{0.4\linewidth}
%    \includegraphics[width=\linewidth]{hist.pdf}
%    \caption{}
%  \end{subfigure}
%  \caption{My flowers.}
%\end{figure}
%
%\begin{figure}[h!]
%\centering
%\includegraphics[width=0.4\linewidth]{original_figure.pdf}
%\caption{One example of the diffraction mapping.}
%\label{fig:original_figure}
%\end{figure}

\noindent{\bf Preprocessing.}
To detect weak signal (``sparse spot'') in diffraction image space, we need to effectively remove the background of the image since the change of interest will be tiny bright spot buried in between the bright rings. We develop a set of preprocessing steps that are tailored to the characteristics of the diffraction images.  
The most important part for the analysis is to remove the largest visible ring (neither the bright area near the center nor the dark area near the boundary of the image). %Therefore, we need highlight this ring. 
This task is nontrivial for the following two reasons. First, there is a dark shadow of a irregularly shaped stick in the middle of the image so the center of the rings is hidden. Second, even if we find the center and remove the shadow we still need identify the the ring with the largest radius. 

\begin{figure}[h!]
\centering
\includegraphics[width=1\linewidth]{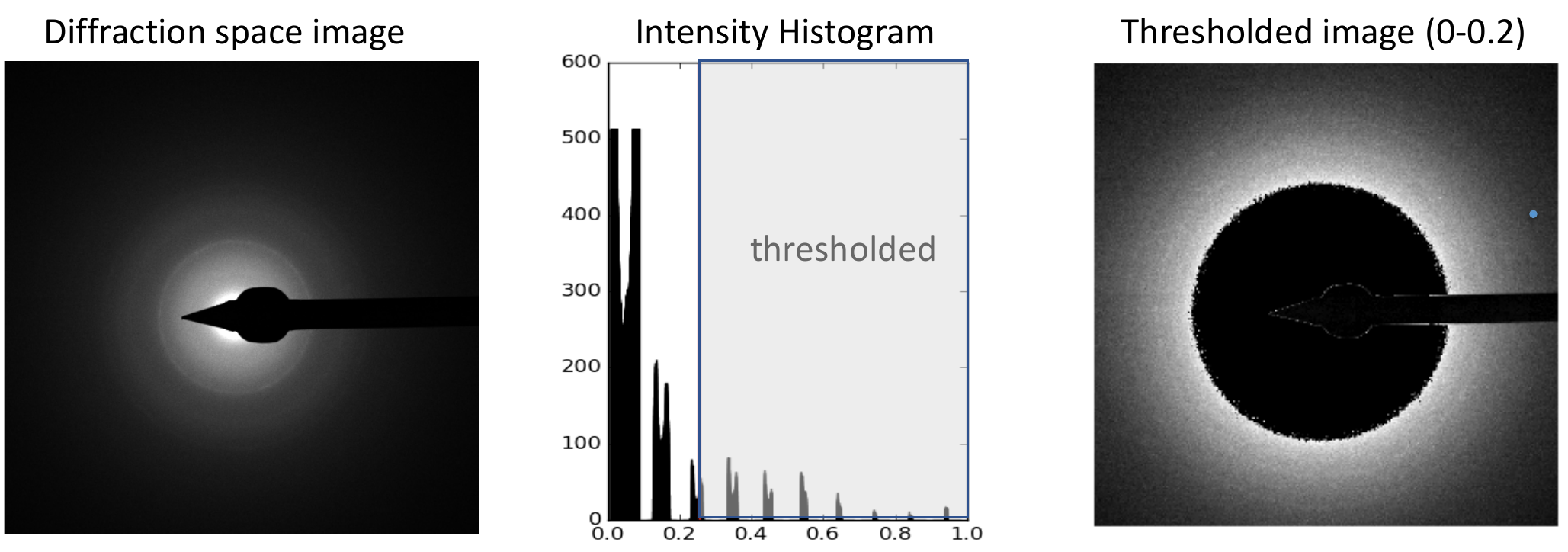}
\caption{Left: a diffraction domain image; Middle: the histogram of the intensity; Right: The thresholded image if we only keep pixels of value in (0, 0.2).}
\label{fig:hist}
\vspace{-0.2in}
\end{figure}

\begin{figure}[h!]
\centering
\includegraphics[width=1.0\linewidth]{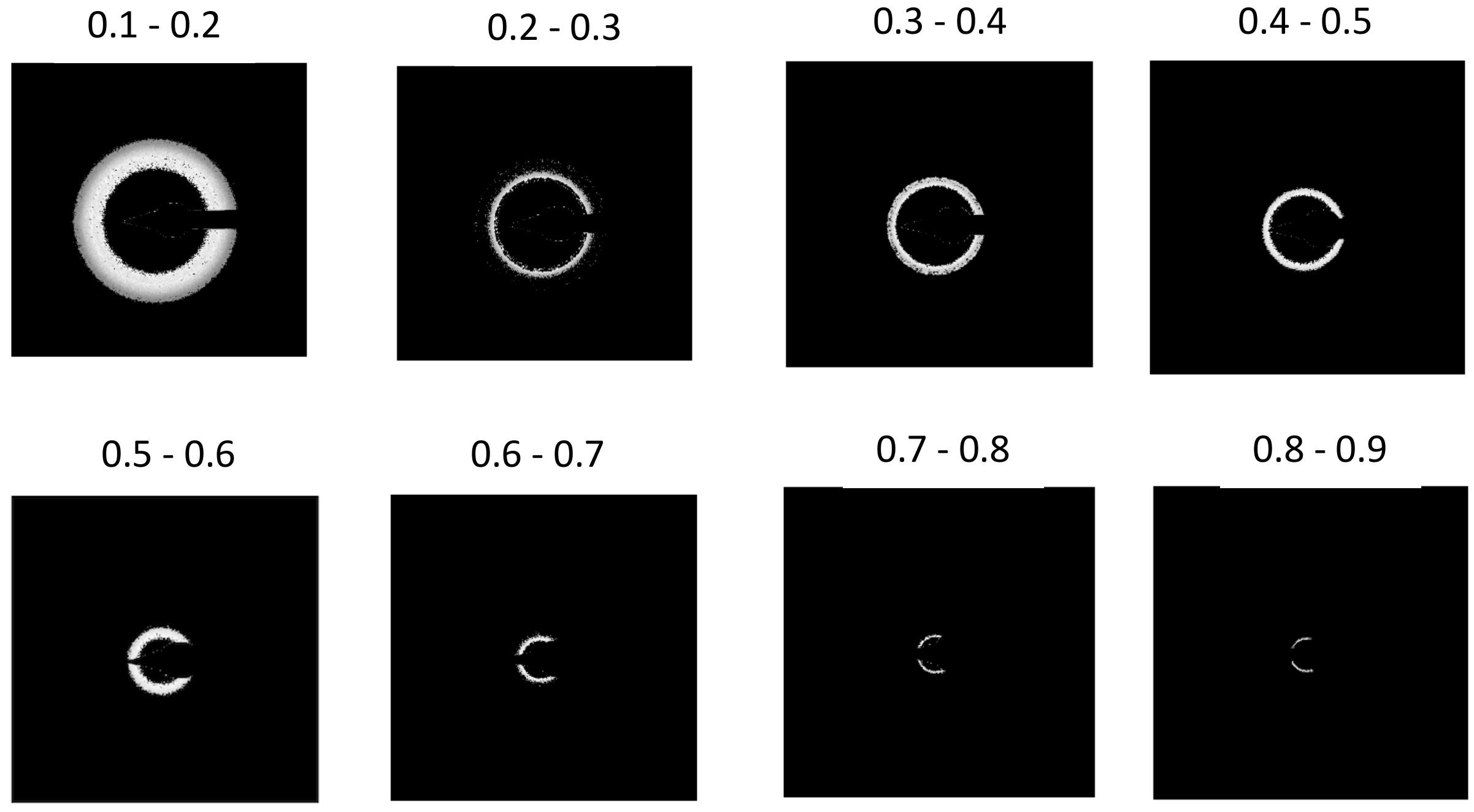}
\caption{Background removal: we threshold a diffraction space image with different range of threshold values, and this yields rings at different radii. These concentric rings help to estimate their common center, and subsequently we subtract off the bright rights to remove these bright rings.}
\label{fig:hist_conti}
\vspace{-0.1in}
\end{figure}

To overcome the first difficulty, we draw the histogram of the pixel values as shown in the left figure in Fig. \ref{fig:hist}.  Several gaps between the brightness is observed in the histogram. For example, there is no point in the image with the brightness centered around $0.4, 0.6$ and $0.8$. In fact, these gaps are highly related to the rings in the image. The right plot in Fig. \ref{fig:hist} shows the points with brightness between $0$ and $0.1$ and we surprisingly find the shadow. Therefore, we can throw away the points with that brightness in order to remove the stick in the original image. Then, Fig. \ref{fig:hist_conti} shows that we can separate the rings successfully by focusing on the points with separated ranges of the brightness. 

\begin{figure}[h!]
\centering
\includegraphics[width=1.0\linewidth]{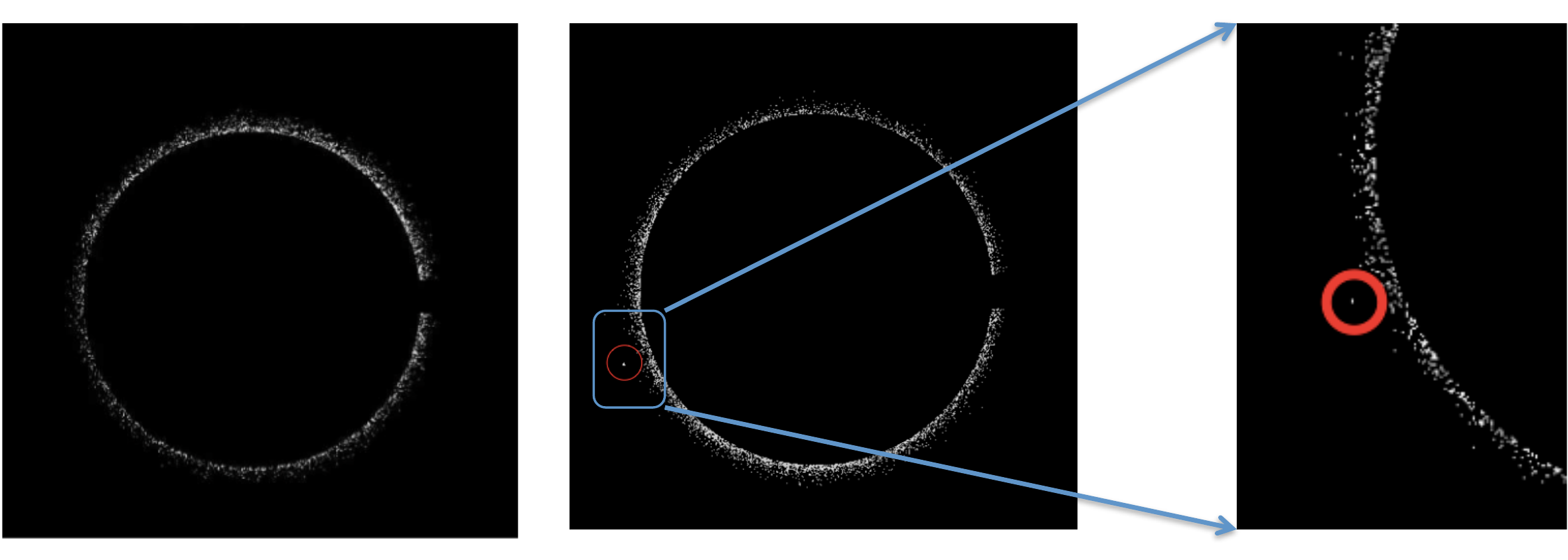}
\caption{The diffraction image after preprocessing for background removal. The complete video is available at www.isye.gatech.edu/$\sim$yxie77/diffraction-video.mp4 %Two persistent bright spots appear from Fram 123 to Frame 212. The left image is the $14$th image of the $100$ images. 
The middle image is the $17$th image of the $100$ images. We zoom out the middle image to show the bright spot which represents the anomaly we would like to detect.}
\label{fig:final}
\vspace{-0.2in}
\end{figure}

To overcome the second difficulties, we use the Hough transformation \cite{vc1962method} to look for the centers. We run the Hough transformation on all the plots in Fig. \ref{fig:hist_conti} and then compute the center by averaging the $8$ estimated centers. To find the largest visible rings, we apply the Canny edge detection algorithm \cite{canny1986computational} that identifies the boundary between the dark and bright area accurately. The final results after all the prepossessing procedures are shown in Fig. \ref{fig:final}. Note that the change - a tiny bright spot can finally be revealed.

\begin{figure}[h!]
\centering
\includegraphics[width=0.7\linewidth]{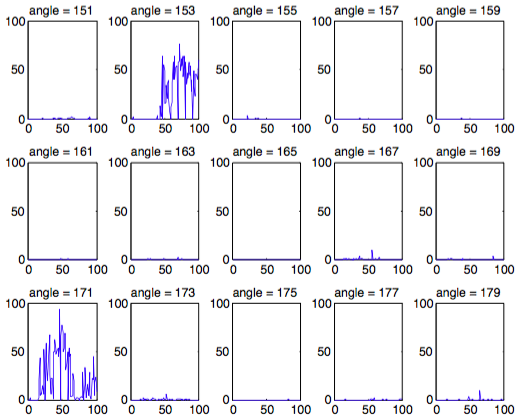}
\caption{The extracted signals for the sequence of $100$ images (selected angles).}
\label{fig:signals}
\vspace{-0.1in}
\end{figure}

The bright spot is very weak and it is even hard to be observed by eyes. Fortunately, we know by domain knowledge that the bright spot usually appear near certain radius (but at an unknown angle). Therefore, we can ``hunt'' the bright spot around such radius. We focus our attention on the points with a specified radius $r$ that is slightly larger than the radius of the ring. We then formulate the detection task as the detection of a sparse mean shift. Using polar coordinate transformation, for each prepossessed image we observe a $360$-dimensional signal that represents the averaged pixel values in every angle for a fixed radius. The results are shown in Fig. \ref{fig:signals}. We can see clearly that one bright spot appears from about the $17$th image for the angle equal to $171$, and another bright spot appears from about the $49$th image for the angle equal to $153$. After the pre-processing, we have a total of $100$ such vectors: $X_1, \ldots, X_{100} \in \mathbb{R}^{d}$, with $d=360$. The pre-change distribution is $\mathcal{N}(0,I_d)$ and the goal is to detect the unknown time $\nu$ at which the underlying distribution changes to $\mathcal{N}(\theta, I_d)$ for some unknown mean $\theta \neq 0$. In this example we just set $\Gamma = \mathbb{R}^d$.

\noindent{\bf Results.} $T_{\rm ACM}(b)$ and $T_{\rm ASR}(b)$ with $\Gamma = \mathbb{R}^d$ both stop at time $t=18$, CUSUM procedure with an all-one post-change mean vector stops at time $t=24$ and GLR procedure stops at time $t=4$. Domain knowledge tells us that the change happens at time $t=17$. So the detection delays of our methods are $1$ while that of CUSUM procedure is $7$ (meaning it takes two samples to detect). The GLR procedure raises an false alarm because it is too sensitive to the noise.

\vfill\pagebreak

\clearpage
\bibliographystyle{IEEEbib}
\bibliography{refs}

\end{document}